\documentclass[%
 reprint,
 superscriptaddress,
 amsmath,amssymb,
 aps,
]{revtex4-1}

\usepackage{graphicx}
\usepackage{dcolumn}
\usepackage{bm}

\begin{document}

\preprint{APS/123-QED}

\title{A new method for measuring excess carrier lifetime in bulk silicon:\\Photoexcited muon spin spectroscopy}

\author{K. Yokoyama}
\email{koji.yokoyama@stfc.ac.uk}
\affiliation{ 
School of Physics and Astronomy, Queen Mary University of London, Mile End, London, E1 4NS, United Kingdom
}%
\affiliation{%
ISIS, STFC Rutherford Appleton Laboratory, Didcot, OX11 0QX, United Kingdom
}%

\author{J. S. Lord}
\affiliation{%
ISIS, STFC Rutherford Appleton Laboratory, Didcot, OX11 0QX, United Kingdom
}%

\author{J. Miao}
\affiliation{ 
School of Physics and Astronomy, Queen Mary University of London, Mile End, London, E1 4NS, United Kingdom
}%
\affiliation{ 
College of Physical Science and Technology, Sichuan University, Chengdu, 610064, People’s Republic of China
}%

\author{P. Murahari}
\affiliation{ 
School of Physics and Astronomy, Queen Mary University of London, Mile End, London, E1 4NS, United Kingdom
}%

\author{A. J. Drew}
\email{a.j.drew@qmul.ac.uk}
\affiliation{ 
School of Physics and Astronomy, Queen Mary University of London, Mile End, London, E1 4NS, United Kingdom
}%
\affiliation{%
ISIS, STFC Rutherford Appleton Laboratory, Didcot, OX11 0QX, United Kingdom
}%
\affiliation{ 
College of Physical Science and Technology, Sichuan University, Chengdu, 610064, People’s Republic of China
}%

\date{\today}

\begin{abstract}
We have measured the optically injected excess carrier lifetime in silicon using photoexcited muon spin spectroscopy. Positive muons implanted deep in a wafer can interact with the excess carriers and directly probe the bulk carrier lifetime whilst minimizing the effect from surface recombination. The method is based on the relaxation rate of muon spin asymmetry, which depends on the excess carrier concentration. The underlying microscopic mechanism has been understood by simulating the four-state muonium model in Si under illumination. We apply the technique to different injection levels and temperatures, and demonstrate its ability for injection- and temperature-dependent lifetime spectroscopy. 
\end{abstract}

\maketitle

Excess carrier lifetime in semiconductors is an extremely sensitive probe of recombination active defect density $N_t$ \cite{Schroder, ReinBook}. In the case of silicon, a lifetime spectroscopy can probe $N_t$ as low as 10$^{10}$ cm$^{-3}$, corresponding to the carrier lifetime in the order of 10 ms. Therefore the lifetime measurements have been utilized to test a quality of Si wafers in various areas, and especially appreciated in photovoltaic applications where the carrier lifetime is a key parameter for the excess carriers to successfully diffuse across the p-n junction in solar cells. The microchip industries have also found its use as a cleanliness monitor in the chip manufacturing processes. It is now widely accepted that there are three main mechanisms that cause the electron-hole pair (EHP) recombination in semiconductors: 1) Shockley-Read-Hall (SRH) recombination (characterized by its lifetime $\tau_{SRH}$), 2) Auger recombination ($\tau_{Auger}$), and 3) radiative recombination ($\tau_{rad}$) \cite{Schroder, ReinBook}. The bulk recombination lifetime $\tau_{bulk}$ is then given by a relation,
\begin{eqnarray}
\tau_{bulk}=\frac{1}{\tau_{SRH}^{-1}+\tau_{Auger}^{-1}+\tau_{rad}^{-1}}
\;
\label{eq:bulk}.
\end{eqnarray}
Among those mechanisms, the SRH recombination is a multiphonon process mediated by deep-level defect centers, and dominates $\tau_{bulk}$ in low-level carrier injections, whilst the Auger recombination plays a key role in high-level injections. The radiative recombination is usually negligible in bulk Si due to the indirect band structure.

Although $\tau_{SRH}$ gives a good indication of the $N_t$ level, it alone cannot determine $N_t$ explicitly  --- it is always necessary to assume the defect type, which is characterized by its energy level and capture cross section for electrons and holes. Deep level transient spectroscopy (DLTS) is therefore commonly utilized to investigate the defect centers \cite{Schroder, ReinBook, Lang}. However Rein et al. \cite{ReinBook, Rein} proposed that injection- and temperature-dependent lifetime spectroscopy (IDLS and TDLS) could provide a direct identification of the defect types. The techniques have been demonstrated for Si samples with intentionally introduced metal impurities \cite{Rein, Rein2}.

To measure the carrier lifetime, there are several traditional and novel methods, such as the photoconductance decay (PCD) and photoluminescence decay measurements. Induction-coupled PCD and its varieties are becoming more popular by virtue of their contactless and non-destructive measurement \cite{Schroder, ReinBook, Cuevas}. These techniques measure, by their nature, the effective lifetime of injected carriers, given by $1/\tau_{eff}=1/\tau_{bulk}+1/\tau_{S}$. The second term represents a contribution from the surface lifetime $\tau_{S}$ which strongly depends on how the wafer surface has been conditioned. It is therefore necessary to extract $\tau_{bulk}$ by 1) treating the surface to make $\tau_{S}$ asymptote either 0 ({\it e.g.} sandblasting) or $\infty$ ({\it e.g.} passivation), or 2) measuring the same samples with different thicknesses $d$ and extrapolating the observed lifetimes for $1/d \rightarrow 0$. Although these methods are established and widely used, there are few experimental techniques to directly measure $\tau_{bulk}$, minimizing uncertainties associated with the surface recombination. Those techniques can be important not only in the semiconductor/photovoltaic material engineering, but also in fundamental understanding of the EHP recombination mechanisms.

In this letter we demonstrate a use of positively charged (anti)muon $\mu^+$ as a contactless probe of $\tau_{bulk}$ in Si. Spin-polarized $\mu^+$ with an energy of 4 MeV (``surface'' muons) are generated in a proton accelerator facility and implanted in bulk materials with the distribution thermalizing over several hundred $\mu$m. In a case of single crystal Si, the implantation depth can be as deep as 700 $\mu$m, where the surface effect is negligibly small in most cases. Our recent upgrade of the HiFi muon spectrometer at the ISIS pulsed neutron and muon source in the UK enables us to photoexcite samples with a high-energy laser pulse \cite{Yokoyama, YokoyamaPS, Wang}. A pulsed muon source is useful for time-differential studies, as well as for achieving a large stimulation by virtue of the high-intensity light source. The sample temperature can be varied in a wide range using cryostats and hot stages available in the HiFi experimental suite \cite{Yokoyama, Lord}. Combining these capabilities, muon spin spectroscopy (collectively known as $\mu$SR, corresponding to muon spin relaxation/rotation/resonance) can not only measure the excess carrier lifetime but also investigate its injection and temperature dependence. The muons are an extremely dilute impurity ($<$10$^5$ cm$^{-3}$) and although the muon centers cause recombination, they should have a negligible effect on the bulk carrier lifetime compared to the other impurities present.

Upon implantation, muons decay with a lifetime of 2.2 $\mu$s and emit positrons preferentially in the muon spin direction, which is then subsequently detected. The obtained time spectrum for muon spin asymmetry carries information on the muon state and its interaction with local atomic/electronic environment \cite{Blundell, Nuccio}. The $\mu$SR technique has been applied to many semiconductor systems, especially to single crystal Si \cite{Patterson, CoxRev}. There have been several $\mu$SR studies on illuminated Si wafers, which report a large photoinduced change in the $\mu$SR time spectrum \cite{KadonoPRL, KadonoPRB03, Fan, ChowProc}.

In semiconductors, an implanted $\mu^+$ can capture an electron to form a muonium atom (Mu = $\mu$$^+$ + $e$$^-$), a radioisotope of hydrogen. As with H, Mu can exist in three charge states in semiconductors: Mu$^0$, Mu$^+$, and Mu$^-$. In addition, in the case of Si, there are two distinct lattice sites for Mu to occupy: the bond-center site (Mu$_{BC}$) and the interstitial tetrahedral site (Mu$_{T}$). The charge state and lattice site depend on the formation energy determined by the dopant type, concentration, and temperature. For instance, the initial muon asymmetry in intrinsic Si in room temperature (RT) consists of nearly equal amount of Mu$_{BC}^+$ and Mu$_{T}^0$ component. The diamagnetic Mu$_{BC}^+$ fraction decreases monotonically as decreasing temperature from 250 K down to 200 K. This behavior is attributed to slowing down of the thermally activated ionization of Mu$_{BC}^0$ into Mu$_{BC}^+$ centers. Therefore, almost the same amount of Mu$_{BC}^0$ and Mu$_{T}^0$ can be found in low temperatures, such as T = 77 K \cite{Patterson, CoxRev}. When light illuminates a Si sample, injected excess carriers start interacting with the Mu centers in a complex mechanism including spin exchange interaction, cyclic charge exchange reaction, and site change reaction \cite{CoxRev, KadonoPRL, KadonoPRB03, Fan, ChowProc, Yokoyama}. These interactions result in a spin relaxation of the bound electron in Mu, which then depolarizes the $\mu ^+$ spin via the hyperfine (HF) interaction. Since the relaxation rate of electron spin is proportional to the excess carrier density $\Delta n$, the muon spin relaxation rate $\lambda$ is sensitive to $\Delta n$, and can be used to measure its dynamics. However this assumption is true only in the low rate regime ({\it i.e.} relaxation rate of electron spin $<$ HF frequency), and the microscopic exchange mechanism is discussed later in this letter.

\begin{figure}
\includegraphics{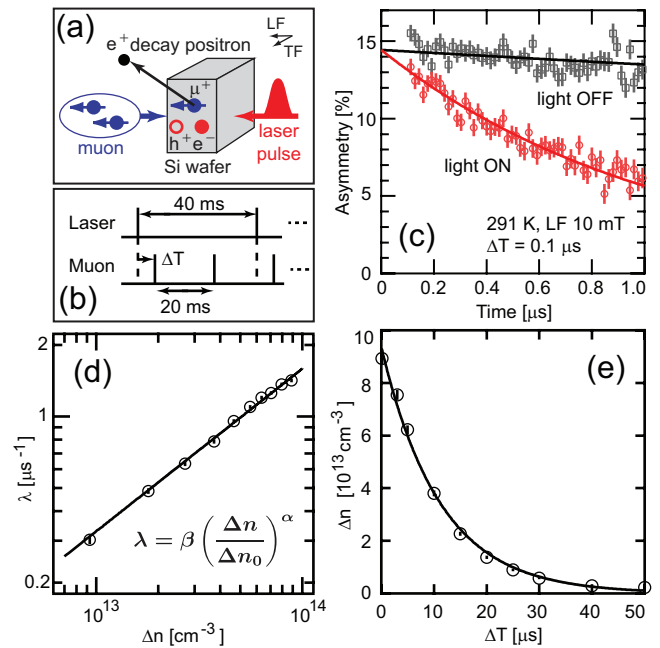}
\caption{\label{fig:291K100G} (Color online) (a) Schematic diagram of the experimental geometry. Magnetic fields are applied either parallel (longitudinal field LF) or perpendicular (transverse field TF) to the direction of muon spin. (b) Timing diagram of laser and muon pulse. Pulse duration (FWHM) of the laser and muon pulse are $\approx$16 and $\approx$70 ns respectively. (c) $\mu$SR time spectra for light OFF (black open squares) and ON (red open circles). 5$\times 10^{6}$ events are averaged for each spectrum. The first 100 ns is removed from the spectra because the good data region is not obtained until the entire muon pulse has arrived at the sample. Fit parameters are $A(0) = 14.44(3)$ \%, $\lambda '$ = 0.068(2) $\mu s^{-1}$ for light OFF, and $\lambda$ = 0.94(2) $\mu s^{-1}$ for light ON. (d) $\lambda$ as a function of $\Delta n$. The data is fitted to a function indicated in the figure, which gives $\alpha = 0.68(4), \; \beta = 1.46(4) \; \mu s^{-1},$ and $\Delta n_0 = 8.9\times 10^{13} cm^{-3}$. (e) Carrier decay curve has been fitted to the single exponential function with $\Delta n(0) = 9.4(4) \times 10^{13} cm^{-3}$ and $\tau = 11.1(9) \; \mu s$.}
\end{figure}
 
Our experiment has been carried out on a 500-$\mu$m thick intrinsic single crystal Si wafer (n-type, R $>$1000 $\Omega$$\cdot$cm, both sides polished) with $\langle$111$\rangle$ axis perpendicular to the surface. As shown in Fig. \ref{fig:291K100G}(a), one side is facing the incoming pump light, whereas the other side faces the muon beam. Details of the experimental setup, including the sample environment, are explained elsewhere \cite{Yokoyama}. The distribution of stopped muon is centered in the wafer by adjusting the number of aluminum foil degraders, with its FWHM estimated to be $\approx$130 $\mu$m by a Monte Carlo simulation. Monochromatic 1064-nm laser light injects excess carriers almost uniformly throughout the sample by virtue of its low absorption in Si. The excess carrier density has been calculated based on an absorption coefficient $\alpha$(293 K) = 14.32 cm$^{-1}$ measured in RT, and $\alpha$(77 K) = 2.37$\times 10^{-2}$ cm$^{-1}$ taken from the literature \cite{Macfarlane}. Because of the long absorption lengths compared with the wafer thickness, we assume that the central density represents $\Delta n$ for the entire sample. The illuminated area on the sample is 9.6 cm$^2$ and covers the entire area of the muon beam. With these geometries and the calculated carrier diffusion lengths ranging 100 -- 200 $\mu$m, the surface effect is negligible in the obtained lifetime spectra. Fig. \ref{fig:291K100G}(b) illustrates the pulse timing, in which muon pulses arrive at the sample at $\Delta T$ after laser pulses. Since the repetition rate of laser and muon are 25 and (pseudo-)50 Hz respectively \cite{Yokoyama}, the muon data are sorted and binned to ``light ON'' and ``light OFF'' spectra, and averaged for statistics, assuming that the photoinduced change is already over after 20 ms. In the optical setup, two attenuator assemblies and calibrated neutral density filters are used to control the pump laser energy accurately for a wide range of carrier injection.

Fig. \ref{fig:291K100G}(c) shows representative light OFF/ON $\mu$SR time spectra for Si in 291 K under LF 10 mT. The initially formed Mu$_{T}^0$ makes a rapid transition to Mu$_{BC}^0$, which is then quickly ionized to Mu$_{BC}^+$. Thus the light OFF time spectrum is constituted of the diamagnetic Mu$_{BC}^+$ center, which shows a very small relaxation because the Zeeman interaction ``locks'' them along the field direction. Upon illumination at  $\Delta T$ = 0.1 $\mu$s generating $\Delta n$ = $4.7\times 10^{13} \;\mathrm{cm}^{-3}$, the muon spin asymmetry shows a significant relaxation. 
Based on the timescale of excess carrier recombination and the microscopic mechanism as described below, we use the first 1 $\mu$s in the spectrum as the fitting range assuming that $\Delta n$ is constant in this period. The light OFF spectrum is fitted to $A(t) = A(0)e^{-\lambda ' t}$ with $A(0)$ and $\lambda '$ as fitting parameters. Then the light ON spectrum is fitted to the same functional form but with fixed $A(0)$, and a free relaxation rate $\lambda$. Because this relaxation rate arises as a consequence of the Mu-photocarrier interaction, we consider that $\lambda$ is a relaxation rate specific for this $\Delta n$. 
$\lambda$ is then measured as a function of $\Delta n$ with fixed $\Delta T$ = 0.1 $\mu s$. The data is then fitted to a power law indicated in Fig. \ref{fig:291K100G}(d) with $\alpha$ and $\beta$ as fit parameters. From the obtained function, it is now possible to calculate $\Delta n$ from a measured $\lambda$. We can therefore measure $\Delta n$ as a function of $\Delta T$ and determine the carrier lifetime. Fig. \ref{fig:291K100G}(e) shows the obtained decay curve and a fit to $\Delta n\left(\Delta T\right) = \Delta n\left(0\right) \mathrm{exp}\left[-\Delta T/\tau\right]$ with $\Delta n\left(0\right)$ and $\tau$ as fit parameters.

Based on the obtained decay constant, $\tau = 11.1 \; \mu s$, which is considered to be equivalent to $\tau_{bulk}$, let us calculate $N_t$ assuming that the defect type is interstitial iron (Fe$_i$), a common deep-level defect center in Si wafers. Because of the moderate injection level, the SRH process is the predominant recombination mechanism {\it i.e.} $\tau_{bulk}\simeq\tau_{SRH}$. The SRH model \cite{Schroder, ReinBook} enables us to calculate $N_t$ based on the recombination parameters of Fe$_i$ in Si in RT: $\Delta E = 0.38 \; \mathrm{eV}, \sigma_n = 5\times 10^{-14} \;\mathrm{cm}^{2}, \; \mathrm{and} \; \sigma_p = 7\times 10^{-17} \;\mathrm{cm}^{2}$, where $\Delta E$ is energy level of the Fe$_i$ defect center measured from the valence band maximum, and $\sigma_n$ and $\sigma_p$ are its capture cross section for electrons and holes respectively \cite{Macdonald}. A straightforward calculation gives $N_t = 6.8\times 10^{13} \;\mathrm{cm}^{-3}$ for this wafer.

\begin{figure}
\includegraphics{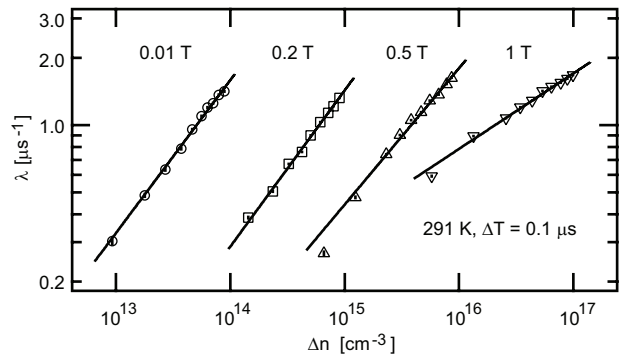}
\caption{\label{fig:291KPscans} Light ON relaxation rate $\lambda$ measured as a function of $\Delta n$ under LF 10 mT (open circles, identical to Fig. \ref{fig:291K100G}(d)), 0.2 T (open squares), 0.5 T (open triangles), and 1.0 T (open inverted triangles). Solid lines denote curves fitted to $\lambda = \beta \left(\Delta n/\Delta n_0\right)^\alpha$. Fit parameters for 0.2, 0.5, and 1.0 T are ($\alpha$, $\beta \; \left[\mu s^{-1}\right]$, $\Delta n_0 \; \left[cm^{-3}\right]$) = (0.70(3), 1.33(3), $9.0\times 10^{14}$), (0.61(6), 1.64(6), $8.6\times 10^{15}$), and (0.34(3), 1.70(4), $9.9\times 10^{16}$) respectively.}
\end{figure}

This lifetime spectroscopy utilizing the $\mu^+$ spin relaxation rate as a probe of $\Delta$n enables us to investigate $\tau_{bulk}$ in a wide range of injection level, an essential parameter for the IDLS measurement, by changing the magnitude of LF. For example, in LF 10 mT, the fit quality for light ON spectra becomes gradually worse when $\Delta n$ exceeds $1 \times 10^{14} \;\mathrm{cm}^{-3}$ because the relaxation rate is too fast (see Fig. \ref{fig:291K100G}(d)). On the other hand if $\Delta n$ is less than $1 \times 10^{13} \;\mathrm{cm}^{-3}$, the fit quality is also poor because the relaxation is now too slow. It is however possible to change the ``sensitivity'' of interaction between the Mu centers and excess carriers by changing the magnitude of LF --- this is corresponding to varying the Zeeman interaction of muon spin with respect to the Mu HF interaction \cite{Chow}. In other words a high field decouples the Mu HF interaction so that the $\mu ^+$ spin is less sensitive to the interaction between Mu centers and injected electrons/holes. Therefore we can tune $\lambda$  for the best fit quality depending on the injection levels. As shown in Fig. \ref{fig:291KPscans}, three more injection levels have been measured for the $\lambda \; vs. \; \Delta n$ curve under 0.2, 0.5, and 1.0 T. The decay curve for each field is measured in the same way as Fig. \ref{fig:291K100G}(e), and gives $\tau$ = 9.4 $\pm$ 0.3, 9.0 $\pm$ 1.1, and 9.2 $\pm$ 0.8 $\mu$s for 0.2, 0.5, and 1.0 T respectively. Based on the same argument as the 10 mT data, $\tau\simeq\tau_{SRH}$. The SRH model with the calculated defect density predicts that $\tau_{SRH}$ is the fastest recombination process and dominates $\tau_{bulk}$ up to $\Delta n \sim 10^{17} \;\mathrm{cm}^{-3}$, which agrees with the obtained lifetimes.

\begin{figure}
\includegraphics{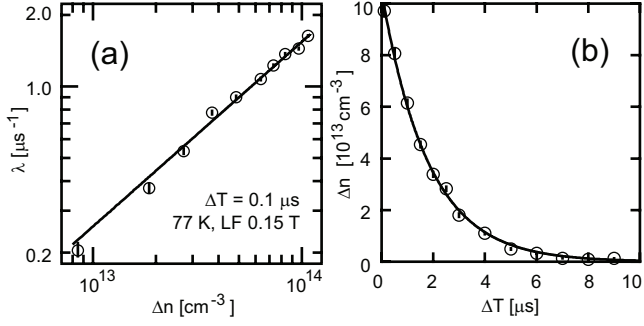}
\caption{\label{fig:77K1500G} Carrier lifetime measurement in 77 K, LF 0.15 T. (a) $\lambda \; vs. \; \Delta n$ curve. The solid line denotes a curve fitted to $\lambda = \beta \left(\Delta n/\Delta n_0\right)^\alpha$, and gives $\alpha = 0.78(7), \; \beta = 1.63(6) \; \mu s^{-1},$ and $\Delta n_0 = 1.1\times 10^{14} cm^{-3}$. (b) Excess carrier decay curve has been fitted to the single exponential with $\Delta n(0) = 1.04(3) \times 10^{14} cm^{-3}$ and $\tau = 1.8(1) \; \mu s$.}
\end{figure}

We now apply the method in a low temperature, 77 K, to demonstrate its feasibility for the temperature dependent measurements. Thermally activated transition from the predominant Mu centers, Mu$_{T}^0$ and Mu$_{BC}^0$, is negligibly small  in this temperature. Under low LF fields, the $\mu$SR signal relaxes fast because of the mobile Mu$_{T}^0$ (see below). As we have seen in Fig. \ref{fig:291K100G}(c) and (d), $\lambda '$ sets the minimum $\lambda$ usable in the $\lambda \; vs. \; \Delta n$ curve. Therefore, for 77 K data, the LF magnitude has been changed to 0.15 T so that $\lambda '$ is negligibly small (0.0118(3) $\mu s^{-1}$), whilst $\Delta n$ is in the same injection level as Fig. \ref{fig:291K100G}(d). Fig. \ref{fig:77K1500G}(a) and (b) show the obtained $\lambda \; vs. \; \Delta n$ and carrier decay curve. The same analysis and fitting method as the 291 K data can be applied and give $\tau = 1.8 \pm 0.1 \; \mu s$. This significantly shorter carrier lifetime is associated with an increase of the capture cross section of defect centers. Considering that the sample temperature is too high for the cascade capture process to be predominant \cite{Lax}, the most likely mechanism is the excitonic Auger recombination \cite{HangleiterTheo}.

The lifetime measurement in low temperature can also be performed using the precession signal of Mu$_{T}^0$, which is readily observable under a weak TF. The same procedure is applied to the Mu$_{T}^0$ signal under TF 0.2 mT to measure the $\lambda \; vs. \; \Delta n$ curve, and subsequently the lifetime spectrum. The difference here is that the Mu$_{T}^0$ precession in dark is already damped ($\lambda ' = 1.50(3) \;\mu s^{-1}$) because of magnetic field inhomogeneity of the instrument and the quantum diffusion of Mu$_{T}^0$, which interacts with impurities in material \cite{Patterson}. Therefore the fit function for light ON should have a decay term, $e^{-(\lambda ' + \lambda) t}$, to find the photoinduced rate separately (see Supplemental Material for details of this analysis \cite{SM}). The obtained carrier lifetime, $\tau = 1.3 \pm 0.3 \;\mu s$, agrees well with the LF measurement (Fig. \ref{fig:77K1500G}), implying that both methods can observe the same excess carrier recombination. However the LF measurement is considered best suited for IDLS and TDLS measurements because of the ability to tune the field for an interested injection level as seen in Fig. \ref{fig:291KPscans}, and the applicability for a wide temperature range. The latter advantage is also endorsed by previous photoexcited $\mu$SR studies on Si, which found a large photoinduced relaxation in the Mu$_{BC}$ precession not only in the low temperature range continuously down to several Kelvins \cite{Fan} but also in the high temperatures up to 550 K \cite{ChowProc}.

\begin{figure}
\includegraphics{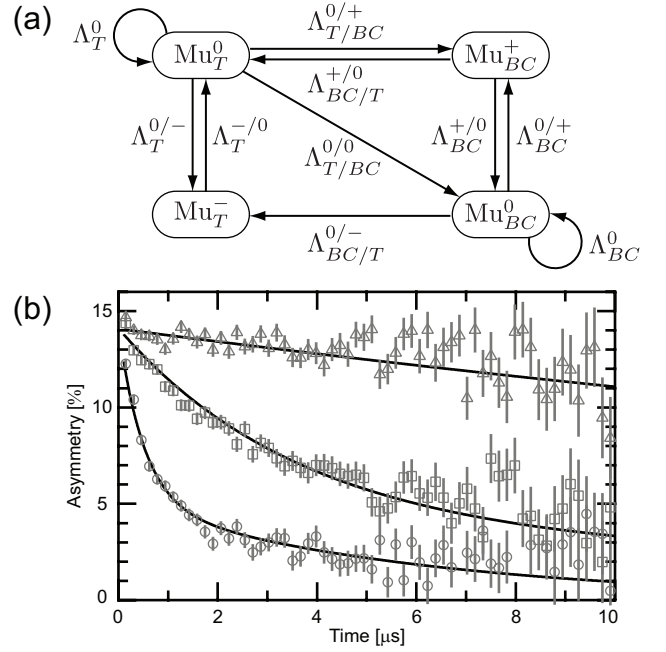}
\caption{\label{fig:Simulation}  (a) The four-state model of Mu in Si under illumination. The notation of $\Lambda$ follows the convention in Ref. \cite{Fan}. Its superscript and subscript indicate the charge-state and site change respectively with a slash between before and after the transition. $\Lambda_T^{0}$ and $\Lambda_{BC}^{0}$ indicate spin exchange interaction in Mu$_T^0$ and Mu$_{BC}^0$ with conduction electrons. (b) Representative light ON ($\Delta$T = 0.1 $\mu$s) $\mu$SR time spectra in 291 K under LF 10 mT for $\Delta$n(0) = 1.0$\times$ 10$^{14}$ $cm^{-3}$ (circles), 1.6$\times$ 10$^{13}$ cm$^{-3}$ (squares), and 1.4$\times$ 10$^{12}$ cm$^{-3}$ (triangles). The solid lines denote the fit (see the main text).}
\end{figure}

\begin{figure}
\includegraphics{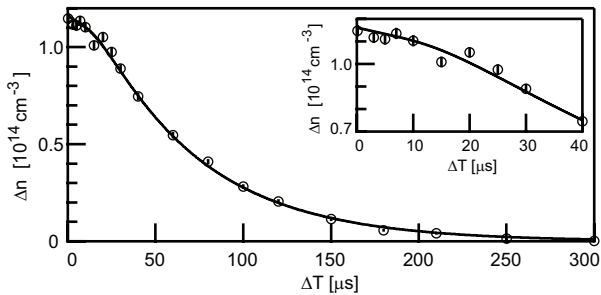}
\caption{\label{fig:LifetimeSpectrum} Carrier decay curve for a 1000-$\mu$m thick intrinsic Si wafer (R $>$10000 $\Omega$$\cdot$cm) in 291 K. Absorption coefficient $\alpha$ measured in RT = 6.80 cm$^{-1}$. The solid line denotes a fit described in the main text. The fit results are: $D$ = 12(2) cm$^2$/s, $\tau_{bulk}$ = 2(1) $\times$ 10$^2$ $\mu$s, and $\Delta n_c$ = 1.16(1) $\times$ 10$^{14}$ cm$^{-3}$. (inset) Magnified view shows the nearly constant $\Delta$n for $\Delta$T $<$ 10 $\mu$s.}
\end{figure}

We have so far demonstrated the lifetime measurements based on the empirical observation that $\lambda$ can be a useful yardstick of $\Delta n$. But what is the underlying microscopic mechanism? To answer this question it is necessary to study the model of Mu dynamics, which was originally used to analyze RF-$\mu$SR data \cite{Kreitzman}, and later applied to a photoexcited $\mu$SR experiment \cite{Fan}. The four-state model shown in Fig. \ref{fig:Simulation}(a) is based on the three-state model used by Fan et al., but has the Mu$_T^-$ state in addition, which becomes more important for $\Delta$n $>$ 10$^{14}$ cm$^{-3}$ \cite{Fan}. Transition from one Mu state to another is characterized by a transition rate $\Lambda$, which can depend on the capture cross section $\sigma$ for electron/hole, electron/hole density, activation energy, and prefactor. Because this network is activated upon photocarrier injection, what we observe in the $\mu$SR spectrum is the dynamics of Mu transition, rather than a signal from static Mu states. To gain the comprehensive picture it is crucial to study a full $\mu$SR time spectrum, where $\Delta$n stays constant throughout. We therefore ran the same set of experiments as Fig. \ref{fig:291K100G} but using a thicker wafer with a longer carrier lifetime to satisfy this condition (see inset in Fig. \ref{fig:LifetimeSpectrum}). Fig. \ref{fig:Simulation}(b) shows three representative $\mu$SR spectra out of eight $\Delta$n's. We perform a simultaneous fit for the spectra with $\sigma$'s as global fit parameters, from which $\Lambda$'s are calculated based on the known $\Delta$n. This simulation and fit have been carried out using QUANTUM \cite{LordQuantum}, a program to solve the time evolution of the muon spin using the density matrix method, running on Mantid \cite{Mantid}. See Supplemental Material for details of this computation \cite{SM}.
The fit result firstly tells us enhanced rates in $\Lambda_{BC}^{0/+}$ and $\Lambda_{BC}^{+/0}$ upon carrier injection. Because the rates are much faster than the HF frequencies in Mu$_{BC}^0$ ($<$92 MHz), the muon spin is hardly depolarized in this cycling transition. Secondly a high $\Delta$n opens two channels, which ``leaks'' the Mu$_{BC}$ states to the others. Some of Mu$_{BC}^+$ escape to the Mu$_T^0$ state via $\Lambda_{BC/T}^{+/0}$, where the HF interaction (2 GHz) depolarizes the muon spin --- this is the fast relaxing part in Fig. \ref{fig:Simulation}(b). Others are converted from Mu$_{BC}^0$ to Mu$_{T}^-$ via $\Lambda_{BC/T}^{0/-}$, which is an inert, less interactive Mu center with carriers --- and this is the subsequent slow relaxing tail. As $\Delta$n decreases, these channels become narrower, resulting in isolated Mu$_{BC}$. Slow decay in the lowest $\Delta$n in Fig. \ref{fig:Simulation}(b) is thus attributed to $\Lambda_{BC}^{0}$ causing the relaxation during the fleeting window when the cycling Mu$_{BC}$ state is in Mu$_{BC}^{0}$. Based on these observations we conclude that the first 1 $\mu$s window used in the lifetime measurement is corresponding to fitting primarily the fast Mu$_{T}^0$ relaxation. Tuning the Mu sensitivity in a higher field (Fig. \ref{fig:291KPscans}) is equivalent to decoupling the HF interaction in Mu$_{T}^0$. The sub-linear dependence of $\lambda$ on $\Delta$n ({\it i.e.} $\alpha$ $\approx$ 0.7) stems from the transition path for Mu$_{BC}^0$ leading to Mu$_{T}^-$.

Finally Fig. \ref{fig:LifetimeSpectrum} shows the carrier decay curve for the long-lifetime wafer. The curve is apparently different from the single exponential decay, but has a shoulder around 20 $\mu$s, where the fast surface recombination driven by carrier diffusion comes into play. This lifetime spectrum can be modelled with a simple 1-dimensional diffusion equation for $\Delta n(z,t)$, $D \frac{\partial^{2}\Delta n}{\partial z^{2}}-\frac{\Delta n}{\tau_{bulk}}=\frac{\partial \Delta n}{\partial t}$, where $D$ is the carrier diffusion constant. Because the wafer surfaces have been lapped and chemically polished, the surface velocity should be $>$10$^4$ cm/s. We therefore assume a boundary condition on the surfaces, $\Delta n(0,t) = \Delta n(d,t) = 0$, and analytically solve the equation with an initial condition, $\Delta n(z,0) = \Delta n_c$. With $D$, $\tau_{bulk}$, and $\Delta n_c$ as fit parameters, the solid line in Fig. \ref{fig:LifetimeSpectrum} shows a fit to $\Delta n(d/2,t)$. The obtained $\tau_{bulk}$ = (2 $\pm$ 1) $\times$ 10$^2$ $\mu$s agrees with $\tau_{bulk} \approx 1 \times 10^2$ $\mu$s, which has been measured by  the wafer manufacturer (PI-KEM Ltd.) using the standard PCD method.

In conclusion, excess carrier lifetime in Si has been measured using photoexcited muon spin spectroscopy. This novel technique enables us to measure $\tau_{bulk}$ directly by virtue of the implanted muons as a bulk probe, and can access a wider range of recombination lifetime (from 50 ns to $>$20 ms), injection level, and temperature. The four-state model has been utilized to explain the underlying microscopic mechanism about how $\lambda$ exhibits the dependence on $\Delta$n. 
The high time resolution, much shorter than the $\tau_{bulk}$, is possible only with a short-pulsed laser, and distinguishes this study from the previous experiments using lamps. The precisely controlled $\Delta$n is achievable with the collimated beam with narrow linewidth, which gives a predictable uniform absorption profile in the sample.
The method can be applied immediately to other semiconductors, such as Ge and GaAs, where interaction of Mu with photoinduced carriers has already been reported \cite{FanGe, YokoyamaGaAs}. Its capability on measuring recombination kinetics can be useful in emerging high-efficiency light harvesting materials.

This work has been supported by European Research Council (Proposal No 307593 - MuSES). We wish to acknowledge the assistance of a number of technical and support staff in ISIS and Queen Mary Univ. of London.


\end{document}